# Security Incident Recognition and Reporting (SIRR): An Industrial Perspective

*Full Paper*

**George Grispos**
Lero – The Irish Software Research Centre
University of Limerick, Ireland
george.grispos@lero.ie

**William Bradley Glisson**
School of Computing
University of South Alabama, USA
bglisson@southalabama.edu

**David Bourrie**
School of Computing
University of South Alabama, USA
dbourrie@southalabama.edu

**Tim Storer**
School of Computing Science
University of Glasgow, UK
timothy.storer@glasgow.ac.uk

**Stacy Miller**
School of Computing
University of South Alabama, USA
sam1522@jagmail.southalabama.edu

## Abstract[1]

Reports and press releases highlight that security incidents continue to plague organizations. While researchers and practitioners' alike endeavor to identify and implement realistic security solutions to prevent incidents from occurring, the ability to initially identify a security incident is paramount when researching a security incident lifecycle. Hence, this research investigates the ability of employees in a Global Fortune 500 financial organization, through internal electronic surveys, to recognize and report security incidents to pursue a more holistic security posture. The research contribution is an initial insight into security incident perceptions by employees in the financial sector as well as serving as an initial guide for future security incident recognition and reporting initiatives.

---

## 1. Introduction

In today's digital societies, responding to security incidents is becoming increasingly imperative in business environments. A Ponemon (2016) study on data breaches reports that 48% of attacks involved malicious activity, 25% were due to negligent human factors, and 27% involved business and information technology process failures. The report goes on to indicate that the mean time to identify an incident is, approximately, 201 days and the mean time to contain an incident once discovered is 70 days. The reality is that the effects of a breach can be very destructive to an organization. This destruction can be experienced in the form of ransomware, system downtime, intellectual property theft, reduced customer confidence, and facilitating attacks on other organizations. The growing concern and impact of ransomware activity, alone, is easily visible in recent news articles (Mogg 2017; Weisbaum 2017).

The continued introduction of new technology, like cloud computing environments (Cahyani et al. 2016; Grispos et al. 2014a; Kynigos et al. 2016), and the necessity to keep policies, standards and procedures up-to-date, as well as fit for purpose continues to make this challenging from a corporate perspective (Grispos et al. 2013). This is further complicated for corporations by the continued introduction and escalating impact of residual data in legal environments (Berman et al. 2015; McMillan et al. 2013). The PricewaterhouseCoopers (2017) report on the Global State of Information Security highlights a growing interest by organizations in threat intelligence. The report goes on to indicate that threat intelligence efforts will become increasingly predictive in nature in an attempt to impede incidents before they transpire. However, researchers and industry professionals alike have pointed out over the years that the weakest link in the security is often the human element, not technology (Berr 2015; Mitnick 2002).

To help combat the deficiencies that are often associated with human interactions, a popular slogan in the broader security arena has become 'if you see something, say something' (Homeland Security 2017; O'Haver 2016). However, in order to resolve and/or minimize a potential security breach, employees in an organization first need to recognize that they have encountered a problem and then successfully report the problem to the appropriate personnel (Mitropoulos et al. 2006). These requirements prompted the idea that organizations are at risk due to a lack of initial security incident cognition, long-term recognition, and internal reporting practice discrepancies. Hence, this research presents the results of an exploratory survey conducted in a Fortune 500 financial organization. The study investigates the ability of individuals within the studied organization to recognize incidents and probes reporting practices. The research contribution is an initial empirical report that indicates issues when it comes to recognizing a security incident along with discrepancies in approaches to reporting. The results of the research provide a foundation for continued research into practical solutions for recognizing, reporting, and resolving security incidents.

The paper is structured as follows. The following section recognizes relevant previous work and the third section presents the research methodology. The fourth section discusses the findings along with an analysis of the results. The final section draws conclusions and identifies future work.

## 2. Relevant Literature

As the number of security incidents affecting organizations continues to increase (Ponemon 2016), it is understandable that these organizations examine different security incident response approaches. Typically, the objective of these incident response approaches is to minimize the damage from an incident, and to allow an organization to ultimately learn about the cause of the incident and how it can be prevented in the future (Grispos et al. 2014b; Mitropoulos et al. 2006). However, before an organization can investigate and learn from an incident, it must be detected through automated approaches (e.g. an Intrusion Detection System) or manually reported by an employee who has noticed something that gives cause for concern (International Organization for Standardization and the International Electrotechnical Commission 2011). Hence, many incident response approaches include a phase (known as the 'detection' or 'identification' phase) to ensure that organizations have the ability to detect, identify, or report the existence of a security incident, which could warrant further investigation (Grispos 2016).

While existing security incident response approaches stress the importance of incident detection and reporting, researchers (Ahmad et al. 2012; Furnell et al. 2007) have speculated that many security incidents go undetected or unreported within organizations, or may be subjected to delays. This is evident in the incident reporting deficiencies identified within national Computer Emergency Response Teams (CERTs) (Koivunen 2010) and organizations with regard to security incident reporting (Ahmad et al. 2012; Grispos et al. 2015; Hove et al. 2014; Metzger et al. 2011; Werlinger et al. 2010). Koivunen (2010) studied six incidents reported to the Finnish national CERT with the purpose of investigating how incidents are detected and reported to the CERT. Koivunen (2010) observed differences in incident response standard solutions for reporting security incidents and the means by which this takes place in reality. Koivunen (2010) identified four main problems from the study including: incident response teams neglecting incident reports from outside their organization, problems identifying the correct individual to whom to report an incident, organizations not exploiting automated incident reporting, and the need for more research to enhance the standards literature so that it reflects real-world requirements.

In addition to examining national CERTs, researchers have also investigated security incident reporting within organizations. Ahmad, et al. (2012) reported that although management, in their studied organization, declared a willingness to investigate reported incidents, there were a number of factors that discouraged employees from reporting such incidents. These include reputational impact, financial penalties and onerous follow-up procedures applied by regulators as a consequence of incidents (Ahmad et al. 2012). Grispos, et al (2015) identified from their case study that the majority of security incidents were reported either verbally or via informal emails, usually to an associate. This is a finding that is shared by Metzgner et al. (2011), who reported that even when automated monitoring mechanisms were in place within their studied organization, the majority of incidents were manually reported through local systems and service administrators, by either telephone or email. Similar results were also observed by Hove, et al. (2014), who concluded that manual reporting mechanisms were more popular within organizations than automatic detection systems. Werlinger, et al. (2010) suggests that this could be because such detection systems lack accuracy and result in a high number of false positives.

While information security researchers have observed problems with security incident reporting within organizations, other researchers (Gonzalez 2005; Reed-Mohn 2007; Schneier 2011) in the safety-critical domain have examined how these problems could be addressed. Schneier (2011) and Reed-Mohn (2007) compared current practices in information security reporting systems with those in the healthcare, aviation, and rail industries and concluded that the quality of practices in information security reporting systems did not match those of their safety-critical equivalents. Gonzalez (2005) has examined the successful implementation of incident reporting programs in aviation and then explored how an equivalent could be constructed for information security. Gonzalez (2005) concluded that one area where information security needs to increase resource investment is security incident learning which would require redirecting efforts from eradication and recovery to quality improvement (Gonzalez 2005). Johnson (2002) examined incident reporting in the rail and healthcare industries and proposed ten barriers that must be addressed before incident reporting can be successfully applied in other industries, including information security. Three of the ten barriers directly relevant to incident recognition and reporting include removing fear of retribution for reporters and 'whistleblowers', encouraging an environment to share information about incidents that could involve friends and colleagues, as well as isolating the fear of media publicity that could arise due to information about an incident becoming public (Johnson 2002). While previous work has examined the importance of security incident reporting and how it could be enhanced by using concepts from other industries, minimal research investigates an organization's ability to recognize and report security incidents.

## 3. Methodology

This research is an initial investigation into an employee's ability to recognize a security incident in a real- world context along with investigating practical reporting tendencies. To investigate these issues, an exploratory survey was conducted in a Global Fortune 500 financial organization. The research methodology is examined from the perspectives of questionnaire development and survey implementation.

### 3.1 Questionnaire Development

The high-level purpose of the survey was trifold. First, it investigates whether individuals within two business units with differing degrees of technology focus could identify an information security incident. Second, it examines how individuals within these business units would report an information security incident and to which individual or group they would report it to within the organization. Third, it inspects the responses from participants to ascertain company policy compliance. In order to obtain this information and nurture an environment that encourages discussing, potentially, commercially sensitive information, the organization's name has been withheld, participating business unit's names have been altered, and any specific information that could identify the organization has been obscured.

The survey questions were developed by the researchers and vetted twice by the organization. The first round of vetting involved two information security managers with an average of 23.9 years of experience. For the last five years, one of these managers was responsible for the information security division within the organization. As a result of the initial round of vetting, one of the questions was completely reworded to improve clarity. In addition, two supplementary correct and wrong options were added to the second question in the survey. The survey was then implemented in a web-based intranet system used by the organization. The second round of vetting utilized five members of the information security team to test the functionality of the web survey tool. No modifications were suggested or implemented as a result of this vetting round. The questions and answer options that were employed in the survey instrument are presented in Table 1 – Survey Questions. The bolded notation in the table indicates the correct answer for the individual questions.

| | |
|---|---|
| 1. | Which business unit do you belong to? (Ops Tech or CM) |
| 2. | Which of the following would you consider to be an information security incident? |
| | • Your laptop or personal computer will not power on, reboots, crashes, locks up or does not respond to your commands. (Yes, **No**, or Don't Know) |
| | • Your organization-issued laptop or mobile device has been lost or stolen. (**Yes**, No, or Don't Know) |
| | • You have noticed a laptop or personal computer on a desk, which has been left unattended for several days/weeks. (**Yes**, No, or Don't Know) |
| | • You notice an unknown individual walking around your department who is not wearing an organizational identity pass. (**Yes**, No, or Don't Know) |
| | • The firewall or antivirus software on your laptop or personal computer is notifying you that your system is being attacked. (**Yes**, No, or Don't Know) |
| | • Your laptop or personal computer is notifying you that your hard disk is full. (Yes, **No**, or Don't Know) |
| | • You can view personal information about people other than yourself that you do not think you should be able to see. (**Yes**, No, or Don't Know) |
| | • You notice that an application you use frequently has become slow and sluggish. (Yes, **No**, or Don't Know) |
| | • You just realized that you sent information, which should have been encrypted, without any encryption. (**Yes**, No, or Don't Know) |
| | • Documents you send to the printer are not printing correctly. (Yes, **No**, or Don't Know) |
| 3. | Have you ever personally experienced an information security incident within the organization within the past decade? (Yes or No) |
| | • If Yes, how many information security incidents have you experienced in the past decade? |
| 4. | Which of the following actions would you first undertake upon noticing that the antivirus software on your laptop or personal computer has notified you that a virus has been detected? |
| | • Email important files you want to save from being affected to a colleague. |
| | • Change your Windows password. |
| | • Run an anti-virus scan and attempt to fix the problem. Contact the IT Helpdesk |
| | • **Report the information security incident.** |
| | • Use Google to find a solution to your problem. |
| | • Remove the network cable from the personal computer or laptop. Turn off the affected personal computer or laptop. |
| 5. | Who would you first contact in the event of you noticing an information security incident? |
| | • Your Line Manager |
| | • IT Helpdesk Services |
| | • Global Desktop Services |
| | • **Information Security** |
| | • The Police |
| | • The Product Vendor |
| 6. | What method would you use to report an information security incident? |
| | • Email the Information Security inbox |
| | • Speak to your team or people leader |
| | • Email your line manager |
| | • **Telephone the information security operations manager** |
| | • Telephone Helpdesk Services |

**Table 1. Survey Questions**

*3.2 Survey Implementation*

The two units that were selected for this experiment, within the organization, include the Operations Technology (Ops Tech) unit and Customer Marketing (CM) business unit. Employees within these two units had undertaken the company's information security training. Normally, this takes place during the employee's induction, where they are presented with the company's information security framework, policies, standards, and guidelines. During induction, security documentation exposure includes documents that affect all employees (e.g. clear desk policy) and those that are job-specific. In addition, employees are also exposed to in-house online videos, which are used to highlight changes in existing security documentation. The survey was advertised via an internal email list and hosted on a local web-based intranet system in October of 2014. The email list contained 1,474 individuals from both the Ops Tech and CM units. A total of 668 participants completed the survey. However, this survey did not include any attention filters or 'trap' questions to filter out individuals that may have been speeding through the survey. Therefore, employee responses that answered all the items for question 2 as 'Yes' (n= 53), 'No' (n=15), or 'Don't know' (n=2) were excluded from the analysis. The final adjusted response rate was 40.57% with 598 participants.

## 4. Results and Analysis

Overall, 485 or 81.10% of respondents were from the Ops Tech unit and 113 or 18.90% were from the CM unit. The number of responses by unit were consistent with the proportional number of employees in each of these units. Table 2 presents the results of the second survey question, which investigates an employee's ability to identify potential incidents. The bolded notation in all of the tables indicates correct answers based on the organization's policy.

To compare the two groups of potential security incident responses a 2x3 Fisher-Freeman-Halton Exact Test (Freeman and Halton 1951) was executed. The results confirmed that the responses for questions that were not a potential security incident were less frequent than expected by chance ($P < .00001$). According to company policies, the first four potential security incidents in Table 2 are not deemed information security incidents. Both units identified the following responses as non-security incidents: receiving a notification that your hard drive is full, an application becoming slow and sluggish, and documents being sent to a printer that are not printed correctly. Surprisingly, 30.72% of the Ops Tech unit and 24.78% of the CM unit identified a laptop/PC not powering on, rebooting, crashing, locking up or does not respond to your commands as a security incident. Overall, 9.48% to 18.76% of employees in each unit did not know whether these four items were a security incident or not.

According to company policies, the final six potential security incidents in Table 2 are deemed information security incidents. Overwhelmingly, both the Ops Tech and CM units were able to identify the following scenarios as information security incidents: lost or stolen organization-issued laptop or mobile device; noticing an unattended laptop/PC; unknown individuals walking around without an organizational identity pass, alerts from firewall or antivirus software notifying you of an attack; the ability to view personal information about other people that you think you should not have access to; and realizing that you just sent information that should have been encrypted but forgot to encrypt the file. The results show that the CM unit was 0.59% to 1.27% more likely than the Ops Tech unit to consider four of these six scenarios were not an actual security incident. In addition, the Ops Tech unit was 0.15% to 1.68% more likely than the CM unit to not know whether five out of the six potential security incident scenarios was actually an incident. The only potential scenario where the CM unit had a slightly higher percentage of employees that did not know if it was a security incident was when firewall or antivirus software on your computer notified you that you system was being attacked (1.77% versus 1.24%). The results indicate that the degree to which a business unit is technologically focused is not a predictor of the ability to identify security incidents.

| Potential Security Incidents | Response | Ops Tech | | CM | | Total | |
|---|---|---|---|---|---|---|---|
| | | Count | % | Count | % | Count | % |
| Your laptop/PC will not power on, reboots, crashes, locks up or does not respond to your commands | Yes | 149 | 30.72% | 28 | 24.78% | 177 | 29.60% |
| | **No** | **270** | **55.67%** | **65** | **57.52%** | **335** | **56.02%** |
| | Don't Know | 66 | 13.61% | 20 | 17.70% | 86 | 14.38% |
| Your laptop/PC is notifying you that your hard disk is full | Yes | 57 | 11.75% | 11 | 9.73% | 68 | 11.37% |
| | **No** | **369** | **76.08%** | **87** | **76.99%** | **456** | **76.25%** |
| | Don't Know | 59 | 12.16% | 15 | 13.27% | 74 | 12.37% |
| You notice that an application you use frequently has become slow and sluggish | Yes | 62 | 12.78% | 17 | 15.04% | 79 | 13.21% |
| | **No** | **332** | **68.45%** | **77** | **68.14%** | **409** | **68.39%** |
| | Don't Know | 91 | 18.76% | 19 | 16.81% | 110 | 18.39% |
| Documents you send to the printer are not printed correctly | Yes | 37 | 7.63% | 8 | 7.08% | 45 | 7.53% |
| | **No** | **402** | **82.89%** | **87** | **76.99%** | **489** | **81.77%** |
| | Don't Know | 46 | 9.48% | 18 | 15.93% | 64 | 10.70% |
| Your organization-issued laptop or mobile device has been lost or stolen | **Yes** | **471** | **97.11%** | **109** | **96.46%** | **580** | **96.99%** |
| | No | 9 | 1.86% | 3 | 2.65% | 12 | 2.01% |
| | Don't Know | 5 | 1.03% | 1 | 0.88% | 6 | 1.00% |
| You have noticed a laptop/PC on a desk which has been left unattended for several days/weeks | **Yes** | **421** | **86.80%** | **100** | **88.50%** | **521** | **87.12%** |
| | No | 43 | 8.87% | 10 | 8.85% | 53 | 8.86% |
| | Don't Know | 21 | 4.33% | 3 | 2.65% | 24 | 4.01% |
| You notice an unknown individual walking around your department who is not wearing an organizational identity pass | **Yes** | **441** | **90.93%** | **104** | **92.04%** | **545** | **91.14%** |
| | No | 38 | 7.84% | 8 | 7.08% | 46 | 7.69% |
| | Don't Know | 6 | 1.24% | 1 | 0.88% | 7 | 1.17% |
| The firewall or antivirus software on your laptop/PC is notifying you that your system is being attacked. | **Yes** | **469** | **96.70%** | **108** | **95.58%** | **577** | **96.49%** |
| | No | 10 | 2.06% | 3 | 2.65% | 13 | 2.17% |
| | Don't Know | 6 | 1.24% | 2 | 1.77% | 8 | 1.34% |
| You can view personal information about people other than yourself that you do not think you should be able to see | **Yes** | **462** | **95.26%** | **108** | **95.58%** | **570** | **95.32%** |
| | No | 13 | 2.68% | 4 | 3.54% | 17 | 2.84% |
| | Don't Know | 10 | 2.06% | 1 | 0.88% | 11 | 1.84% |
| You just realized that you sent information which should have been encrypted without any encryption | **Yes** | **465** | **95.88%** | **108** | **95.58%** | **573** | **95.82%** |
| | No | 11 | 2.27% | 4 | 3.54% | 15 | 2.51% |
| | Don't Know | 9 | 1.86% | 1 | 0.88% | 10 | 1.67% |

**Table 2. Responses to Potential Security Incidents**

When the survey results are compared, in conjunction with the stated corporate policies, only 33.43% of Ops Tech unit and 31.86% of the CM unit got all ten of the options correct in Table 2. The data supports the idea that security incident recognition is an issue that needs to be addressed. Possible solutions to the problem could be training, educational opportunities and technology-based prompts within the organization from a security incident perceptive. The number of correct answers for the second question by department are provided in Table 3.

| Number of Correct Answers | Ops Tech Count | Ops Tech % | CM Count | CM % | Total Count | Total % |
|---|---|---|---|---|---|---|
| 0 | 0 | 0.00% | 0 | 0.00% | 0 | 0.00% |
| 1 | 2 | 0.41% | 0 | 0.00% | 2 | 0.33% |
| 2 | 1 | 0.21% | 1 | 0.88% | 2 | 0.33% |
| 3 | 7 | 1.44% | 2 | 1.77% | 9 | 1.51% |
| 4 | 2 | 0.41% | 0 | 0.00% | 2 | 0.33% |
| 5 | 7 | 1.44% | 4 | 3.54% | 11 | 1.84% |
| 6 | 41 | 8.45% | 7 | 6.19% | 48 | 8.03% |
| 7 | 57 | 11.75% | 12 | 10.62% | 69 | 11.54% |
| 8 | 79 | 16.29% | 17 | 15.04% | 96 | 16.05% |
| 9 | 127 | 26.19% | 34 | 30.09% | 161 | 26.92% |
| 10 | 162 | 33.40% | 36 | 31.86% | 198 | 33.11% |
| Total | 485 | 100% | 113 | 100% | 598 | 100% |

Table 3. Number of Correct Answers

When asked about personal experience with information security breaches in the past decade, the majority of the respondents indicated that they had not experienced a breach. Overall, both units had 23.08% of their employee's indicating that they had a breach experience. When these results are considered in conjunction with the previous questions, opportunities were highlighted in the area of education and recognition. Are the perceived breaches really breaches? On the other hand, are employees who indicated that they did not experience a breach, missing breaches due to their inability to recognize security issues? The responses to breach experience are presented in Table 4. Since this survey did not ask employees about their length of employment in each of the business units and a decade is too wide a range of time for most employees to accurately account for the number of security breaches that they had personally been involved in, it is possible that the responses may be skewed by employees that had worked in units for a longer period of time. Based on this reasoning, the average number of incidents is not reported. In hindsight, it would have been better to use 'in the past year' to get more accurate responses for the exact number of security that employees had been involved.

| Response | Ops Tech Count | Ops Tech % | CM Count | CM % | Total Count | Total % |
|---|---|---|---|---|---|---|
| Yes | 111 | 22.89% | 27 | 23.89% | 138 | 23.08% |
| No | 330 | 68.04% | 81 | 71.68% | 411 | 68.73% |
| Don't Know | 44 | 9.07% | 5 | 4.42% | 49 | 8.19% |
| Total | 485 | 100% | 113 | 100% | 598 | 100% |

Table 4. Breach Experience

The responses to the fourth question, inquiring about actions first undertaken upon notification that a virus has been detected, are presented in Table 5. The top three responses were to contact the IT helpdesk, remove the network cable, and report the information security incident. The correct answer, according to the organization, is to take no action and report the incident. Only 18.56% of the Ops Tech group and 15.93% of the CM group got this question correct. All of the other answers provided, potentially, interfere with an incident response investigation. For example, running a virus scan or turning off a computer may remove residual data from the device depending on the software configuration settings. The helpdesk representative could inadvertently delete residual data or change existing software configurations that would complicate an investigation. Changing the password and removing the network cable could complicate access to the machine from an investigation perspective. This data also indicates that the Ops Tech unit is twice as likely to remove the network cable as opposed to the CM unit.

|  | Ops Tech | | CM | | Total | |
| --- | --- | --- | --- | --- | --- | --- |
| **Response** | **Count** | **%** | **Count** | **%** | **Count** | **%** |
| Change your Windows password | 1 | 0.21% | 0 | 0.00% | 1 | 0.17% |
| Run an anti-virus scan and attempt to fix the problem | 28 | 5.77% | 3 | 2.65% | 31 | 5.18% |
| Contact the IT Helpdesk | 231 | 47.63% | 72 | 63.72% | 303 | 50.67% |
| **Report the information security incident** | **90** | **18.56%** | **18** | **15.93%** | **108** | **18.06%** |
| Use Google to find a solution to your problem | 0 | 0.00% | 0 | 0.00% | 0 | 0.00% |
| Remove the network cable from the personal computer or laptop | 111 | 22.89% | 14 | 12.39% | 125 | 20.90% |
| Turn off the affected personal computer or laptop | 24 | 4.95% | 6 | 5.31% | 30 | 5.02% |
| Total | 485 | 100% | 113 | 100% | 598 | 100% |

**Table 5. Virus Notification**

The responses to the question inquiring as to who should be contacted in the event that an information security incident is noticed are summarized in Table 6. The correct answer according to stated policy is to contact the information security group prior to taking any action. The results indicate that there are behavioral differences between the two business units. The majority of individuals in both groups indicated that they would contact their line manager before contacting the information security group. In the Ops Tech group, the second most popular option was to contact the information security group followed by the helpdesk. In the CM group, the second and third options were reversed. These responses indicate that both departments need to be retrained since only 20.57% of the organization has a strong grasp of the current breach policies especially when that pertains to who should be contacted in the event of an information security incident.

|  | Ops Tech | | CM | | Total | |
| --- | --- | --- | --- | --- | --- | --- |
| **Response** | **Count** | **%** | **Count** | **%** | **Count** | **%** |
| Line Manager | 301 | 62.06% | 73 | 64.60% | 374 | 62.54% |
| IT Helpdesk Services | 72 | 14.85% | 26 | 23.01% | 98 | 16.39% |
| Global Desktop Services | 2 | 0.41% | 1 | 0.88% | 3 | 0.50% |
| **Information Security** | **110** | **22.68%** | **13** | **11.50%** | **123** | **20.57%** |
| The Police | 0 | 0% | 0 | 0.00% | 0 | 0.00% |
| Total | 485 | 100% | 113 | 100% | 598 | 100% |

**Table 6. Contact**

The responses to the query about the preferred method for communicating a security incident are summarized in Table 7. The correct answer according to company policy is to notify the information security manager by telephone. However, the results of this survey indicate that the culturally favored method for communicating a security incident is to speak with a team leader. By reporting potential security incidents to team and people leaders, employees are inadvertently increasing the time that this report takes to reach the correct individual, in this case the information security manager. As a result, the organization's mean-time to identify, respond and resolve a potential security incident is increased and, in turn, could result in increased damages to systems which potentially translates into larger financial losses.

|  | Ops Tech | | CM | | Total | |
| --- | --- | --- | --- | --- | --- | --- |
| Response | Count | % | Count | % | Count | % |
| Email the Information Security inbox | 54 | 11.13% | 10 | 8.85% | 64 | 10.70% |
| Speak to your team or people leader | 254 | 52.37% | 57 | 50.44% | 311 | 52.01% |
| Email your line manager | 9 | 1.86% | 2 | 1.77% | 11 | 1.84% |
| **Phone the Info. Security manager** | **62** | 12.78% | **12** | 10.62% | **74** | 12.37% |
| Telephone the IT Helpdesk | 106 | 21.86% | 32 | 28.32% | 138 | 23.08% |
| Total | 485 | 100% | 113 | 100% | 598 | 100% |

Table 7. Method of Communication

## 5. Conclusions

As technology continues to integrate into all aspects of corporate interactions, responding to security incidents is a vital activity in today's organizations. Hence, being able to recognize and effectively report an incident is paramount in todays' digital atmosphere. This initial investigation into an organization's conceptual understanding of a security incident and the applied reporting practices solicited 598 respondents out of, potentially, 1474 participants from two units in a Global Fortune 500 financial organization. The response rate for this study was 40.57%.

While the results indicated that individuals within both the technology-focused and non-technology-focused business units would take similar actions when a security incident occurs these actions were not in compliance with the studied organization's policies. In addition, the results of the Fisher-Freeman-Halton Exact Test indicate that there is a discrepancy on what actually constituted a security incident in the organization. Hence, there is an opportunity to improve information security incident recognition and reporting within both business units by focusing education initiatives on activities that will provide individuals with specific information on exactly 'what to do' and 'when to do it' when they detect or identify an information security incident. The investigation of technology-oriented reminders, as a vehicle for improving employee compliance and decreasing incident occurrence, is a viable possibility for future research. Future research should also investigate the effectiveness of scenario-based training modules to help employees recognize incidents.

From a broader perspective, the survey also highlights practical divergences in stated corporate policy, like contacting the information security team in the event of a breach and the real-world propensities, like employees contacting the help desk when they perceive a security incident problem. This raises a wider debate concerning the need to align corporate policies, standards and procedures with real-world organizational habits. Should organizations modify policies, standards and procedures to coincide with employee cultural tendencies or should organizations attempt to modify their cultures to comply with stated policies? In either case, this initial study highlights the need to propagate this information effectively and efficiently in an organization along with validation of real-world actions. Future research will investigate survey improvements and implementations in a variety of industries in order to compare results, identify potential trends and formulate possible solutions.

## Acknowledgements

This work was partially supported by SFI Grant No. 13/RC/2094 and ERC Adv Grant. No. 291652 (ASAP).